\documentclass[12pt]{article}
\usepackage{amsfonts}

\begin{document}

\title{Deformed Heisenberg algebra and minimal length }

\author{T. Mas{\l}owski$^1$\footnote{T.Maslowski@proton.if.uz.zgora.pl},
A. Nowicki$^1$\footnote{A.Nowicki@if.uz.zgora.pl}, V. M. Tkachuk$^2$\footnote{tkachuk@ktf.franko.lviv.ua}\\
$^1$Institute of Physics, University of Zielona G\'ora \\
Prof. Z. Szafrana 4a, 65-516 Zielona G\'ora, Poland \\
$^2$Department of Theoretical Physics,\\ Ivan Franko National
University of Lviv,\\ 12 Drahomanov St., Lviv, UA-79005, Ukraine}

\maketitle

\begin{abstract}
A one-dimensional deformed Heisenberg algebra $[X,P]=if(P)$ is studied.
We answer the question: For what
function of deformation $f(P)$ there exists a
nonzero minimal uncertainty in position (minimal length).
We also find an explicit expression for the minimal length in the case of arbitrary function of deformation.
\end{abstract}

\section{Introduction}
Recently much attention has been devoted to the study of deformed Heisenberg
algebras of different kinds. In this paper we focus on
deformed algebras with minimal length. History of this subject is
very long. Snyder's paper \cite{Snyder47} was the first
publication on this subject. In that paper the Lorentz-covariant
deformed Heisenberg algebra leading to quantized
space-time was proposed. For a long time there were only a few paper on this
subject. The interest to deformed algebras was renewed after
investigations in string theory and quantum gravity which suggest the
existence of a nonzero minimal uncertainty in position following
from the generalized uncertainty principle (GUP). In
\cite{Kem95,Kem96} it was shown that GUP and nonzero minimal
uncertainty in position can be obtained from a modified Heisenberg
algebra, where in the right hand side of it a term proportional to
squared momentum is added. Subsequently there were published many
papers where different quantum system in space with deformed
Heisenberg algebra was studied. They are one-dimensional harmonic
oscillator with minimal uncertainty in position \cite{Kem95} and
also with minimal uncertainty in position and momentum
\cite{Tkachuk1,Tkachuk2}, $D$-dimensional isotropic harmonic
oscillator \cite{chang, Dadic}, three-dimensional Dirac oscillator
\cite{quesne} and one-dimensional Coulomb problem \cite{fityo},
(1+1)-dimensional Dirac oscillator with Lorentz-covariant deformed
algebra \cite{Quesne10909}, three-dimensional Coulomb problem with
deformed Heisenberg algebra in the frame of perturbation theory
\cite{Brau,Benczik,mykola,Stet,mykolaOrb}, singular inverse square
potential with a minimal length \cite{Bou1,Bou2}, ultra-cold
neutrons in gravitational field with minimal length
\cite{Bra06,Noz10,Ped11}, composite system in deformed space with
minimal length \cite{Quesne10,Bui10}.

In this paper we study a general deformation of one-dimensional
Heisenberg algebra, when the right hand side of it is some function of
momentum. As we know, up to now there has been no answer to the question
about the existence of a minimal length in this general case. The
aim of this paper is to fill this gap.

\section{Minimal length}

We consider a modified one-dimensional Heisenberg algebra generated by position $\mathbf{X}$
and momentum $\mathbf{P}$ hermitian operators satisfying
\begin{eqnarray} \label{a1}
[\mathbf{X}\,,\mathbf{P} ] \ = \ if(\mathbf{P}),
\end{eqnarray}
where $f$ is called function of deformation and we assume that it is strictly positive ($f > 0$),
even function (cf. \cite{Kem95}).

In momentum representation both operators acting on a square
integrable functions $\phi(p) \in \mathcal{L}^2(-a\,,a;f)\, ,(a
\leq \infty)$
\begin{eqnarray}
\mathbf{P}\,\phi(p) &=& p\,\phi(p)\, ,\label{aa1}\\
\mathbf{X}\,\phi(p) &=& i\,f(p) \frac{d}{dp}\,\phi(p)\,
.\label{aa2}
\end{eqnarray}
where the norm of $\phi$ is given by
\begin{equation}
\parallel \phi\parallel^2 \ = \ \int_{-a}^a\,\frac{dp}{f(p)} \mid\phi(p)\mid^2\, .\label{aa3}
\end{equation}
The hermiticity of $\mathbf{X}$ demands $\phi(-a) = \phi(a) = 0$.

The aim of this paper is to answer the question: {\it for what
function of deformation $f(p)$ there exists a
nonzero minimal uncertainty in position} $\Delta_\phi(\mathbf{X}) \geq \Delta(\mathbf{X})_{min}$.
Nonzero minimal uncertainty in position $\Delta(\mathbf{X})_{min} = l_0$ is called nonzero minimal length.

Further we use the following definitions of the mean value
$\langle \mathbf{A}\rangle_\phi$ and dispersion
$\Delta_\phi(\mathbf{A})$ of some operator $\mathbf{A}$ in the
state $\phi \in \mathcal{L}^2(-a\,,a; f)$
\begin{eqnarray}
\langle \mathbf{A} \rangle_\phi &=& \int_{-a}^a\,\frac{dp}{f(p)}\,\phi^*(p)\,\mathbf{A}\,\phi(p)\, ,\label{aa4}\\
\Delta^2_\phi(\mathbf{A}) &=& \langle \mathbf{A}^2\rangle_\phi -
\langle \mathbf{A}\rangle^2_\phi \ = \
\int_{-a}^a\,\frac{dp}{f(p)}\,\phi^*(p)\left(\mathbf{A} - \langle
\mathbf{A}\rangle_\phi\right)^2 \phi(p)\, ,\label{aa5}
\end{eqnarray}
for normed states $\parallel\phi\parallel^2 =
\langle\mathbf{I}\rangle_\phi = 1$.

Let us recall two well known facts  which follow from the Heisenberg
uncertainty relation
\begin{eqnarray}
\Delta^2_\phi(\mathbf{X})\,\Delta^2_\phi(\mathbf{P}) \ \geq \
\frac{1}{4} \langle f(\mathbf{P})\rangle^2_\phi\, .\label{un}
\end{eqnarray}
The first one is that for the non-deformed case when $f(p)=1$ the minimal
length is zero. The second one states that in the case of function
of deformation $f(p) =1+\beta p^2$ the minimal length is nonzero and
reads \cite{Kem95,Kem96}
\begin{eqnarray}\label{mlK}
l_0 \ = \ \Delta(\mathbf{X})_{min} \ = \ \sqrt\beta.
\end{eqnarray}
For these two cases the momentum $p$ is given on the full line
$-\infty<p<\infty$.

In the case of some other functions of deformation it is also
possible to get minimal length using Heisenberg uncertainty
relation. But in general case of arbitrary $f(p)$ it is difficult
to find minimal length using (\ref{un}) and to give answer about
the existence of minimal length.

The idea of this paper is to relate the deformed algebra characterized by $f(P)$ to one of these two algebras, namely, either to the non-deformed ($f(P)=1$) one or to
the deformed one characterized by $f(P)=1+\beta P^2$.
We find that the minimal length in the first case is zero and in the second is nonzero. Moreover we also
find the value of the minimal length.

One can also consider nonlinearly transformed momentum operator $\mathbf{Q} = h(\mathbf{P})$,
where a function $q = h(p)$ is continuous, strictly increasing on interval $[-a\,,a]$
and operator $\mathbf{X}$ is the same for both algebras. Under this mapping we obtain
a new deformed algebra related to (\ref{a1}) and satisfying the relation
\begin{eqnarray}\label{a2}
[ \mathbf{X}\,,\mathbf{Q} ] \ = \ \,ig(\mathbf{Q})\,,\qquad g(q) \ = \ f(p) \frac{d q}{d p}\, ,
\end{eqnarray}
and we assume that the function of deformation $g$ similarly as $f$ is a positive
even function \footnote{This assumption means that $h(p)$ is an odd function.}.

Using function $q = h(p)$ we can change variable
$\tilde{\phi}(h(p)) = \phi(p)$ and we get realization of
$\mathbf{Q}\,,\mathbf{X}$ in the space of square integrable
functions $\tilde{\phi}(q) \in \mathcal{L}^2(-b\,,b;g)$
\begin{equation}
\mathbf{Q}\,\tilde{\phi}(q) \ = \ q\,\tilde{\phi}(q)\, ,\qquad
\mathbf{X}\,\tilde{\phi}(q) \ = \ i\,g(q)\frac{d \tilde{\phi}}{d
q}\, ,\label{aa7}
\end{equation}
and the norms of state in both spaces are equal
\begin{equation}
\parallel\phi\parallel^2 \ = \ \int_{-a}^a \frac{d p}{f(p)} \mid\phi(p)\mid^2 \
= \ \int_{-b}^b \frac{d q}{g(q)} \mid\tilde{\phi}(q)\mid^2 \ = \
\parallel\tilde{\phi}\parallel^2\, ,\label{aa8}
\end{equation}
which follows from (\ref{a2}). The same holds for dispersions
\begin{equation}
\Delta^2_\phi(\mathbf{A}) = \Delta^2_{\tilde{\phi}}(\mathbf{A})\, .
\end{equation}

From the second equation in (\ref{a2}) we find relation between
$p$ and $q$
\begin{eqnarray}\label{hp}
\int_0^p{d p' \over f(p')}=\int_0^q{d q' \over g(q')}\, ,
\end{eqnarray}
which implicitly defines the transformation $q=h(p)$.
In this case function $h(p)$ maps the domain $-a \leq p \leq a$ onto $-b=h(-a) \leq q \leq h(a) =b$.
When such mapping is possible the minimal length will be the same for two algebras (\ref{a1}) and (\ref{a2}).

From (\ref{hp}) it follows that
\begin{equation}
\int_0^a \frac{d p}{f(p)} \ = \ \int_0^b \frac{d q}{g(q)}\, ,\label{aa6}
\end{equation}
which is equivalent that a mapping $q = h(p)$ from the domain $-a
\leq p \leq a$ to the domain $-b \leq q \leq b$ is possible.

We consider two cases. The first case
\begin{equation}
\int_0^a \frac{d p}{f(p)} \ = \ \infty\, .\label{ab7}
\end{equation}
In order to fulfill this condition we can put $g = 1$ with $q$
given on the full line
\begin{equation}
\int_0^a \frac{d p}{f(p)} \ = \ \int_0^\infty d q \ = \ \infty\, .\label{ab8}
\end{equation}
So, in this case algebra (\ref{a1}) is mapped to non-deformed one ($g = 1$) on the full line and therefore
the minimal lenght is zero.

In the second case
\begin{equation}
\int_{0}^a \frac{d p}{f(p)} \ = \  {\rm const} \ < \ \infty\, .\label{aa9}
\end{equation}
Now in order to fulfil (\ref{aa9}) we can choose $g(q) = 1 + \beta q^2$ and $b=h(a)=\infty$.
Then $\beta$ can be found from the equation
\begin{equation}
\int_0^a \frac{dp}{f(p)} \ = \ \int_0^\infty \frac{dq}{1 + \beta q^2} \ = \ \frac{\pi}{2\sqrt{\beta}}\, .\label{aa10}
\end{equation}
In this case algebra (\ref{a1}) is mapped to the deformed algebra proposed
by Kempf \cite{Kem95,Kem96} and according to (\ref{mlK}) the minimal length is
\begin{eqnarray}\label{ml}
l_0 \ = \ \frac{\pi}{2}\,\left(\int_0^{a}{d p \over
f(p)}\right)^{-1}\, .
\end{eqnarray}

Let us consider more explicitly a few examples.

\noindent{\it Example 1.}
\begin{eqnarray}
f(p) \ = \ e^{\alpha p^2},
\end{eqnarray}
where $-\infty < p < \infty$. This function of deformation in the
case $\alpha>0$ was recently considered in \cite{Dor11}.

For $\alpha = \lambda^2 >0$ using (\ref{ml}) we find that the minimal length is
\begin{eqnarray}\label{E1ml1}
l_0 = \lambda\sqrt\pi.
\end{eqnarray}
In the case $\alpha\le 0$ the minimal length is zero.

For the case of $\alpha = \lambda^2 $  it is also possible to find the
minimal length in another way using the fact that in this case the
function of deformation as function of $p^2$ is convex. As a
result the Heisenberg uncertainty relation (\ref{un}) reads \cite{Dor11}
\begin{eqnarray}\label{unr}
\Delta^2_\phi(\mathbf{X}) \Delta_\phi^2(\mathbf{P}) \ \ge \
{1\over 4}\langle e^{\lambda^2 \mathbf{P}^2}\rangle_\phi^2 \ \ge \ {1\over 4}e^{2 \lambda^2 \langle \mathbf{P}^2\rangle_\phi}\, ,
\end{eqnarray}
and one can find  that
\begin{eqnarray}\label{E1ml2}
l_0=\lambda\sqrt{e \over 2}\, .
\end{eqnarray}
As we see our method gives a better result for minimal length (\ref{E1ml1})
in comparison with (\ref{E1ml2}).

\noindent{\it Example 2.}
\begin{eqnarray}\label{f}
f(p) \ = \ (1+\lambda^2 p^2)^{\alpha}\, ,
\end{eqnarray}
where $-\infty < p < \infty$.

For $\alpha\le 1/2$ the minimal length is zero and in the case
of $\alpha > 1/2$ the minimal length reads
\begin{eqnarray}
l_0 \ = \ {\lambda}\,{\sqrt\pi\,\Gamma(\alpha)\over\Gamma(\alpha-1/2)}\, .
\end{eqnarray}
It is worth to note that in the case of $\alpha > 1$ the function
of deformation as function of $p^2$ is convex and using this fact
it is possible to get some result for minimal length similarly as
in the first example. For other $\alpha$ the function of
deformation is not convex, nevertheless our method gives a
possibility to obtain the result for the minimal length.

\noindent{\it Example 3.} To make Example 2 complete let us consider
\begin{eqnarray}\label{ff}
f(p) \ = \ (1-\lambda^2 p^2)^{\alpha}\, ,
\end{eqnarray}
where $-1/\lambda \leq p \leq 1/\lambda$.

We find that for $\alpha\ge 1$ the minimal length is zero and for $\alpha <1$ we
obtain
\begin{eqnarray}
l_0 \ = \ {\lambda}\,{\sqrt\pi\,\Gamma(3/2-\alpha)\over\Gamma(1-\alpha)}\, .
\end{eqnarray}

\section{Conclusion}

In this paper we have studied deformed algebras (\ref{a1}) with
a symmetric function of deformation and answered
the following question: For what function of deformation
$f(p)$ the minimal length is nonzero? Answer to this question
is given by equation (\ref{ml}) which presents the minimal
length in the case of an arbitrary function of deformation and is the
main result of the paper. When $\int_0^{a}{d p \over f(p)}$ is
finite the minimal length is nonzero and when this integral diverges the
minimal length is zero. Using (\ref{ml}) we can calculate an
explicit expression for the minimal length for different functions
of deformation which is demonstrated in this paper by several examples.

\section*{Acknowledgment}
VMT thanks for warm hospitality the University of Zielona G\'ora where this paper was done.


\begin{thebibliography}{99}
\bibitem{Snyder47} H. S. Snyder, Phys. Rev. {\bf 71}, 38 (1947).
\bibitem{Kem95} A. Kempf, G. Mangano, R. B. Mann,
        Phys. Rev. D {\bf52}, 1108 (1995).
\bibitem{Kem96} A. Kempf,
        Phys. Rev. D {\bf54}, 5174 (1996).
\bibitem{Tkachuk1} C. Quesne and V. M. Tkachuk, J. Phys. A
{\bf 36}, 10373 (2003).
\bibitem{Tkachuk2} C. Quesne and V. M. Tkachuk, J. Phys. A
{\bf 37}, 10095 (2004).
\bibitem{chang} L. N. Chang, D. Minic, N. Okamura and T. Takeuchi,
Phys. Rev. D {\bf 65}, 125027 (2002).
\bibitem{Dadic} I. Dadi\'{c}, L. Jonke and S. Meljanac, Phys. Rev.
D {\bf 67},  087701 (2003).
\bibitem{quesne} C. Quesne and V. M. Tkachuk, J. Phys. A {\bf 38},
  1747 (2005).
\bibitem{fityo} T. V. Fityo, I. O. Vakarchuk and V. M. Tkachuk,
J. Phys. A {\bf 39}, 2143 (2006).
\bibitem{Quesne10909} C. Quesne and V. M. Tkachuk, J. Phys. A {\bf 39},
  10909 (2006).
\bibitem{Brau} F. Brau, J. Phys. A {\bf 32},  7691 (1999).
\bibitem{Benczik} S. Benczik, L. N. Chang, D. Minic and T.
Takeuchi, Phys. Rev. A {\bf 72},  012104 (2005).
\bibitem{mykola} M. M. Stetsko and V. M. Tkachuk, Phys. Rev. A
{\bf 74},  012101 (2006).
\bibitem{Stet}M. M. Stetsko, Phys. Rev. A
{\bf 74},  062105 (2006).
\bibitem{mykolaOrb} M. M. Stetsko and V. M. Tkachuk, Phys. Lett. A
{\bf 372},  5126 (2008).
\bibitem{Bou1}  Djamil Bouaziz, Michel Bawin, Phys.Rev.A {\bf 76}, 032112
(2007).
\bibitem{Bou2}  Djamil Bouaziz, Michel Bawin, Phys.Rev.A {\bf 78}, 032110
(2008).
\bibitem{Bra06}  F. Brau, F. Buisseret, Phys.Rev.D {\bf 74}, 036002
(2006).
\bibitem{Noz10} Kourosh Nozari, Pouria Pedram, EPL 92,
50013 (2010).
\bibitem{Ped11} Pouria Pedram, Kourosh Nozari, S. H. Taheri, JHEP 1103:093,
(2011).
\bibitem{Quesne10}  C. Quesne, V.M. Tkachuk,  Phys. Rev. A {\bf
81}, 012106 (2010).
\bibitem{Bui10}  F. Buisseret, Phys.Rev. A {\bf 82}, 062102
(2010).
\bibitem{Dor11}  Glauber Dorsch, Jose Alexandre Nogueira,
Minimal Length in Quantum Mechanics via Modified Heisenberg
Algebra, arXiv:1106.2737

\end{thebibliography}
\end{document}